\documentclass[%
onecolumn,%
oneside,%
floats,%
aps,%
prd,%
nobibnotes,%
nofootinbib,%
amsmath,%
amssymb,%
amsfonts,%
amscd,%
superscriptaddress,%
eqsecnum%
]{revtex4}

\usepackage[utf8]{inputenc}
\usepackage{graphicx, array, dcolumn}
\usepackage[paperwidth=210mm, paperheight=297mm, centering, hmargin=1.8cm, vmargin=2.5cm]{geometry}

\newcommand{\be}{\begin{equation}}
\newcommand{\ee}{\end{equation}}

\newcommand{\lsim}{\mbox{$<$\hspace{-0.8em}\raisebox{-0.4em}{$\sim$}}}

\def\be{\begin{eqnarray}}
\def\ee{\end{eqnarray}}

\def\-g{\sqrt{-g}}

\renewcommand\rho{\varrho}

\newcommand{\vpc}{\varphi_c}
\newcommand{\vp}{\varphi}

\begin{document}

\title{Shape of the inflaton potential and the efficiency of the universe heating}
\author{A.D.\,Dolgov}
\email{dolgov@fe.infn.it} \affiliation{Novosibirsk State University, Novosibirsk
630090, Russia} \affiliation{ITEP, Bol.\,Cheremushkinskaya ul. 25, Moscow 113259,
Russia}
\affiliation{Dipartimento di Fisica e Scienze della Terra, Universit\`a degli Studi di Ferrara\\
Polo Scientifico e Tecnologico -- Edificio C, Via Saragat 1, Ferrara 44122, Italy}

\author{A.V.\,Popov}
\email{popov@izmiran.ru} \affiliation{Pushkov Institute of Terrestrial Magnetism,
Ionosphere and Radio Wave Propagation,\\ Troitsk, Moscow 142190, Russia}

\author{A.S.\,Rudenko}
\email{a.s.rudenko@inp.nsk.su} \affiliation{Novosibirsk State University,
Novosibirsk 630090, Russia} \affiliation{Budker Institute of Nuclear Physics,
Novosibirsk 630090, Russia}

\begin{abstract}
It is shown that the efficiency of the universe heating by an inflaton field depends
not only on the possible presence of parametric resonance in the production of
scalar particles but also strongly depends on the character of the inflaton approach
to its mechanical equilibrium point. In particular, when the inflaton oscillations
deviate from pure harmonic ones toward a succession of step functions, the
production probability rises by several orders of magnitude. This in turn leads to a
much higher temperature of the universe after the inflaton decay, in comparison to
the harmonic case. An example of the inflaton potential is presented which creates a
proper modification of the evolution of the inflaton toward equilibrium and does not
destroy the nice features of inflation.
\end{abstract}

\maketitle

\section{Introduction}\label{s-intro}
Cosmological inflation consisted, roughly speaking, of two epochs. The first one was
a quasi-exponential expansion, when the Hubble parameter, $H$, slowly changed with
time and the universe expanded by a huge factor, $e^N$, where
\begin{equation}\label{N}
N = \int H dt \gg 1.
\end{equation}
During this period the Hubble parameter exceeded the inflaton mass or, rather, the
square of the Hubble parameter was larger than the second derivative of the inflaton
potential:
\begin{equation} \label{H2-U2}
H^2 > \left|\frac{d^2U (\phi)}{d\phi^2}\right| \equiv \left|U'' (\phi)\right|,
\end{equation}
where $U(\phi)$ is the potential of the inflaton field, $\phi$. Due to the large
Hubble friction (see Eq.~\eqref{eq:ddot-phi}) during this epoch the field $\phi$
remained almost constant, slowly moving in the direction of the "force",
$-U'(\phi)$.

The second stage began when $H^2$ dropped below $\left|U''(\phi)\right|$ and
continued till the inflaton field reached the equilibrium value where
$U'(\phi_{eq})= 0$. It is usually assumed that $\phi_{eq} =0$ and $U(\phi_{eq}) =
0$. The last condition is imposed to avoid a nonzero vacuum energy. During this
period $\phi$ oscillated around $\phi_{eq}$, producing elementary particles, mostly
with masses smaller than the frequency of the inflaton oscillations. This was a
relatively short period which may be called a big bang, when the initial dark
vacuum-like state exploded, creating hot primeval cosmological plasma.

The process of the universe heating was first studied in Refs.~\cite{ad-al, abbott,
turner} within the framework of perturbation theory. A non-perturbative approach was
pioneered in Refs.~\cite{ad-dk, brand-1, brand-2}, where the possibility of
excitation of parametric resonance which might grossly enhance the particle (boson)
production rate was mentioned. In the model of Ref.~\cite{ad-dk} parametric
resonance could not be effectively induced because of the redshift and scattering of
the produced particles which were dragged out of the resonance zone and the main
attention in this work was set on non-perturbative production of fermions. However,
the resonance may be effective if it is sufficiently wide. In this case the particle
production rate can be strongly enhanced~\cite{brand-1, brand-2, kls-1, kls-2}.

As is well known, parametric resonance exists only in the process of the boson
production. In quantum language it can be understood as Bose amplification of
particle production due to the presence of identical bosons in the final state; it
is the same phenomenon as the induced radiation in laser. For bosons there could be
another phenomenon leading to very fast and strong excitation of the bosonic field
coupled to inflaton, if the effective mass squared of such field became negative (a
tachyonic situation)~\cite{tachyon-1, tachyon-2, tachyon-3}. It happens for
sufficiently large and negative product $g \phi$; see below
Eqs.~\eqref{ddot-chi-flat},~\eqref{L-int}. This is similar to the Higgs-like effect,
when the vacuum state becomes unstable. However, in contrast to the Higgs
phenomenon, this took place only during a negative half-wave of the inflaton
oscillations.

Both phenomena are absent in the case of fermion production. The imaginary mass of
the fermions breaks the hermicity of the Lagrangian, so tachyons must be absent. As
for parametric resonance, it is not present in the fermionic equations of motion.
The latter property is attributed to the Fermi exclusion principle. A
non-perturbative study of the fermion production shows that the production
probability sharply grows when the effective mass of the fermions crosses
zero~\cite{ad-dk, tkachev-pp}.

The resonance amplification of the particle production by the inflaton exists not
only in the canonical case of harmonic inflaton oscillations, i.e. for the potential
$m^2\phi^2/2$,  but for a very large class of inflaton potentials. The explicit
magnitude of the production probability depends, of course, upon the shape of the
potential, but there is no big difference between different power law potentials,
$\sim \phi^n$. However, as we show in this work, the production rate could be
drastically enhanced for some special forms of the potential, if the potential
noticeably  deviates from a simple power law.

There are several more phenomena which might have an impact on the particle
production probability. In Refs.~\cite{noise-1, noise-2, noise-3} the effects of
quantum or thermal noise on the inflaton evolution have been studied. It was argued
by these two groups that the resonance is not destroyed by noise. If the noise is
not correlated temporally it would lead on  the average to an increase in the rate
of particle production~\cite{noise-2, noise-3}. This result is valid both for
homogeneous and inhomogeneous noise. However, the cosmological expansion was
neglected in Refs.~\cite{noise-2, noise-3} and it possibly means that the resonance,
to survive in realistic cosmological situations, should be sufficiently wide.

The efficiency of the production depends also on the model of inflation. In
particular, in the case of multifield inflation the canonical parametric resonance
is suppressed~\cite{batt-2008, batt-2009, braden}. However, efficient (pre)heating
is possible via tachyonic effects. To this end a trilinear coupling of the inflaton
to light scalars is necessary. If the tachyonic mechanism is not operative, the old
perturbative approach~\cite{ ad-al, abbott, turner} would be applicable.

A new mechanism of enhanced preheating after multifield inflation has been found in
the recent paper~\cite{batt-3}, due to the presence of extra produced species which
became light in the course of the multifield inflaton evolution.

The efficiency of different earlier scenarios of the cosmological heating after
inflation is discussed in a large number of review papers, e.g.~\cite{heat-rev},
where one can find an extensive list of literature. More recent development is
described in reviews~\cite{pp-rev-1, pp-rev-2}.

In this paper we study a parametric resonance excitation for different forms of the
inflaton potential, $U(\phi)$. A new effect is found: for some non-harmonic
potentials of a single field inflation parametric resonance (not tachyonic) is
excited considerably stronger than in the case of simple power law potentials, both
in flat space-time and in cosmology. Correspondingly, the cosmological particle
production by the end of inflation would be much more efficient, and the temperature
of the created plasma would become noticeably higher. In Sect.~\ref{s-res-stat} we
consider this problem in flat space-time to get a feeling for a proper choice of the
inflaton potential that could generate the signal $\phi(t)$ most efficient for the
particle production. Consideration of the flat space-time example clearly
demonstrates the essence of the effect which is not obscured by the cosmological
expansion. In Sect.~\ref{s-res-exp} we study the evolution of the inflaton field in
cosmological background for different potentials $U(\phi)$. Based on the example
considered in Sect.~\ref{s-res-stat}, we found a potential for which the inflaton
induces parametric resonance much more efficiently than in the purely harmonic case,
or other simple power law potentials. We also comment there on the properties of
inflationary cosmology with such modified inflaton potentials. In this section the
scalar particle production rate by such "un-harmonic" inflaton is calculated. The
results are compared to the particle production rate for the usual harmonic
oscillations of the inflaton. Section~\ref{s-back} is dedicated to an estimate of
the effects of back reaction of the particle production on the inflaton evolution.
In Sect.~\ref{s-conclud} we draw conclusions.

We have chosen the sign and the amplitude of the initial inflaton field to avoid or
to suppress tachyonic amplification of the produced field $\chi$.

\section{Parametric resonance in flat space-time}\label{s-res-stat}

Let us consider at first an excitation of parametric resonance in the classical
situation, when the space-time curvature is not essential and the Fourier amplitude
of  the would-be resonating scalar field $\chi$ satisfies the equation of motion
\begin{equation} \label{ddot-chi-flat}
\ddot \chi + \left(m_\chi^2+ {k^2} + g \phi\right) \chi = 0,
\end{equation}
where $m_\chi$ is the mass of $\chi$, $k$ is its momentum, and $g$ is the coupling
constant between $\chi$ and another scalar field $\phi $ with the interaction
\begin{equation}\label{L-int}
L_{int} = -\frac{1}{2} g \phi \chi^2.
\end{equation}
The classical field $\phi$ is supposed to be homogeneous, $\phi = \phi (t)$ and to
satisfy the equation of motion:
\begin{equation}\label{ddot-phi-0}
\ddot\phi +  U'(\phi) = -\frac{1}{2} g \chi^2.
\end{equation}
Below we mostly neglect the effects of the r.h.s. term in this equation. Its impact
on the inflaton evolution is relatively weak. It introduces  the back reaction of
the particle production and is discussed in Sect.~\ref{s-back}.

Our task here is to determine $U(\phi)$, so that the parametric resonance for $\chi$
would be most efficiently excited. An optimal meander form of the inflaton field
$\phi$ is suggested by the phase parameter approach in the theory of parametric
resonance~\cite{popov}. Here we demonstrate that just a slight shift from the
standard Mathieu model toward an optimal inflaton potential leads to a drastic
increase of the particle production rate. We compare the modified results with the
standard case when the potential of $\phi$ is quadratic, $U(\phi)=m^2 \phi^2/2$, and
therefore Eq.~\eqref{ddot-phi-0} with zero r.h.s. has a solution
$\phi(t)=\phi_0\cos(mt+ \theta)$, where the amplitude $\phi_0$ and the phase
$\theta$ can be found from the initial conditions. One can choose the moment $t=0$
in such a way that $\theta=0$, i.e. $\phi(t)=\phi_0\cos mt$. Substituting this
expression into Eq.~\eqref{ddot-chi-flat}, we come to the well-known Mathieu
equation:
\begin{equation}\label{Mathieu}
\ddot \chi + \omega^2_0\left(1 + h\cos mt\right) \chi = 0,
\end{equation}
where $\omega_0^2=m_\chi^2 + k^2 $ and $h=g \phi_0/\omega^2_0$. When $h\ll 1$ and
the value of $m$ is close to $2\omega_0/n$ (where $n$ is an integer),
Eq.~\eqref{Mathieu} describes a parametric resonance, i.e. field $\chi$ oscillates
with an exponentially growing amplitude. For $h \ll 1$ the solution of
Eq.~\eqref{Mathieu} can be represented as a product of a slowly (but exponentially)
rising amplitude by a quickly oscillating function with frequency $\omega_0$:
\begin{equation}\label{chi-of-t}
\chi = \chi_0 (t) \cos (\omega_0 t + \alpha).
\end{equation}
The amplitude $\chi_0$ satisfies the equation
\begin{equation} \label{xhi0-dot}
-\ddot\chi_0 \cos(\omega_0 t + \alpha) + 2 \omega_0 \dot\chi_0 \sin(\omega_0 t +
\alpha) = h \omega^2_0 \chi_0 \cos mt\,\cos(\omega_0 t +\alpha).
\end{equation}
Let us multiply Eq.~\eqref{xhi0-dot} by  $\sin (\omega_0 t + \alpha)$ and average
over the period of oscillations. The right hand side would not vanish on the
average,
 if $m = 2\omega_0$. In this case $\chi_0$ would exponentially rise if $\alpha = \pi/4$:
\begin{equation}\label{chi0-of-t}
\chi_0 \sim \exp \left(\frac{1}{4} h \omega_0 t\right).
\end{equation}
In this way we recovered the standard  results of the parametric resonance theory.

The rise of the amplitude of $\chi$ is determined by the integral
\begin{equation}\label{energy-chi}
\frac{1}{2}\dot\chi^2 + \frac{1}{2}\omega^2_0 \chi^2 = -g \int dt \chi \dot\chi
\phi\,,
\end{equation}
as one can see from Eq.~\eqref{ddot-chi-flat}. It can be shown that the maximum rate
of the rise is achieved when $\phi$ is a quarter-period meander function (an
oscillating succession of the step-functions with proper step duration); see
Ref.~\cite{popov}.

Let us note that the behavior of  the solution for $\chi$ would dramatically change
with rising $h$. For a large $h$ the eigenfrequency squared of $\chi$ noticeably
changes with time. It may approach zero and, if $|h| >1$, it would even  become
negative for a while; see Eq.~\eqref{Mathieu}. In the latter (tachyonic) case $\chi$
would rise much faster than in the case of classical parametric resonance. We
postpone the study of tachyonic case for a future work, while here we confine
ourselves to a non-tachyonic situation.

\begin{figure}[h]
\center
\includegraphics[scale=0.8]{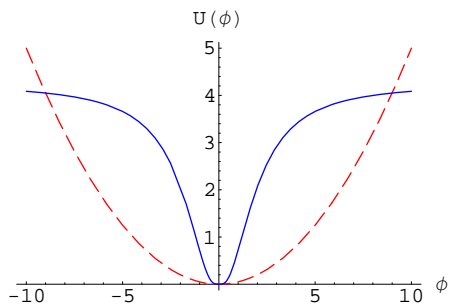}
\caption{\label{fig:U} The potential of the inflaton $U(\phi)$. {\it The red dashed
line} is the quadratic potential $U(\phi) = m^2 \phi^2/2$, and {\it the blue line}
is the potential~\eqref{U-of-phi}. The parameters are $m=10^{-6}m_{Pl}$,
$\lambda_0=85 $, $\lambda_2=4$, $\lambda_4= 1$. The field $\phi$ is measured in
units of $m_{Pl}$ and $U(\phi)$ is measured in units of $10^{-11}\,m_{Pl}^4$}
\end{figure}

To demonstrate an increase of the excitation rate for an anharmonic oscillation we
choose, as a toy model, the potential for the would-be inflaton field $\phi$
satisfying the following conditions: at small $\phi$ it approaches  the usual
harmonic potential, $U \rightarrow m^2 \phi^2 /2$, while for $\phi \rightarrow
\infty$ it tends to  a constant value. It is intuitively clear that in such a
potential field $\phi$ would live for a long time in the flat part of the potential
and quickly change sign near $\phi =0$. This behavior can be rather close to a
periodic succession of the step-functions which we mentioned above. As an example of
such a potential we take
\begin{equation}\label{U-of-phi}
U(\phi) = \frac{1}{2}m^2 \phi^2 \cdot \frac{1+\lambda_0  (\phi/
m_{Pl})^2}{1+\lambda_2 ( \phi/m_{Pl})^2+\lambda_4(\phi/m_{Pl})^4},
\end{equation}
where $m, \lambda_0, \lambda_2, \lambda_4$ are some constant parameters, with $m$
having dimension of mass, the $\lambda_j$ being dimensionless. Here $m_{Pl}$ is the
Planck mass. Observational data on the density perturbations induced by the inflaton
demand $m\simeq 10^{-6}m_{Pl}$ in the model with the inflaton potential $U(\phi) =
m^2 \phi^2/2$, so we use $m=10^{-6}m_{Pl}$ as a reference value throughout the
paper. Since the potential of our toy model is different from the harmonic one, the
inflationary density perturbations could also be different. With the particular
choice of parameters $\lambda_j$, which is presented below, the density
perturbations require a somewhat smaller $m$, approximately by an order of
magnitude. However, our aim here is not the construction of a realistic inflationary
model, but the demonstration of a new phenomenon of more efficient excitation of the
parametric resonance for some forms of inflaton oscillations. To this end we take
$\lambda_0=85$, $\lambda_2=4$, $\lambda_4= 1 $. The plots of the potentials are
shown in Fig.~\ref{fig:U}. Here and below the red (or dashed) curves are for the
quadratic potential $U(\phi) = m^2 \phi^2/2$ and the blue ones are for
potential~\eqref{U-of-phi}. For our choice of $\lambda$ parameters the plots cross
each other at the points $\phi = 0, \pm\,9\,m_{Pl}$.

We did not look for a theoretical justification for the chosen form of the potential
(\ref{U-of-phi}), bearing in mind the large freedom for possible forms of the scalar
field potential at large mass/energy scale, but it is noteworthy that a similar type
of the potential which is quadratic near the minimum and is flatter away from the
minimum was studied in Refs.~\cite{flat-pot-1, flat-pot-2}. Such potentials were in
turn derived in Refs.~\cite{pot-theor-1, pot-theor-2, pot-theor-3, pot-theor-4,
pot-theor-5, axion-pot}. Quoting Ref.~\cite{flat-pot-2}: "This choice was motivated
by monodromy and supergravity models of inflation~\cite{pot-theor-1, pot-theor-2,
pot-theor-3, pot-theor-4, pot-theor-5} and a recent model of axion
quintessence~\cite{axion-pot}".

Figure~\ref{fig:U} for the potential (\ref{U-of-phi}) has a shape which is quite
close to that depicted in Fig.~1 from Ref.~\cite{flat-pot-2}. The properties of the
inflationary model, which might be realized with the inflaton potentials of the type
(\ref{U-of-phi}) can be understood from the results of Refs.~\cite{flat-pot-1,
flat-pot-2}.

\begin{figure}[h]
\center
\includegraphics[scale=0.8]{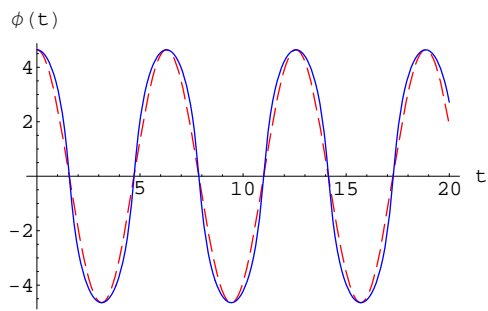}
\caption{\label{fig:phi} Comparison of the numerical solution $\phi_U (t)$ of
Eq.~\eqref{ddot-phi-0} with the potential~\eqref{U-of-phi} ({\it blue line}) and the
harmonic solution $\phi_h(t)=\phi_0\cos mt$ with $\phi_0=4.64$ and $m=10^{-6}m_{Pl}$
({\it red dashed line}). The time $t$ is measured in units of $m^{-1}$ and the field
$\phi$ is measured in units of $m_{Pl}$}
\end{figure}

We solved Eq.~\eqref{ddot-phi-0} with zero r.h.s. and with the
potential~\eqref{U-of-phi} numerically, using Mathematica (here as well as in the
rest of the paper), and compared this solution, $\phi_U (t)$, with the harmonic
solution, $\phi_h(t) =\phi_0\cos mt$. The results are presented in
Fig.~\ref{fig:phi} for the initial conditions $\phi(0)= 4.64\,m_{Pl}$,
$\dot\phi(0)=0.$ The value of the amplitude $\phi_0=4.64$ is chosen because in this
case the frequencies of $\phi_U (t)$ and $\phi_h(t)$ are approximately equal.

For the chosen shape of the potential~\eqref{U-of-phi}, the function $\phi_U (t)$
differs from a cosine toward the step function. Therefore, one can expect that
parametric resonance would be excited stronger for the potential~\eqref{U-of-phi}
than for the quadratic one.

\begin{figure}[h]
\center
\begin{tabular}{c c c}
\includegraphics[scale=0.8]{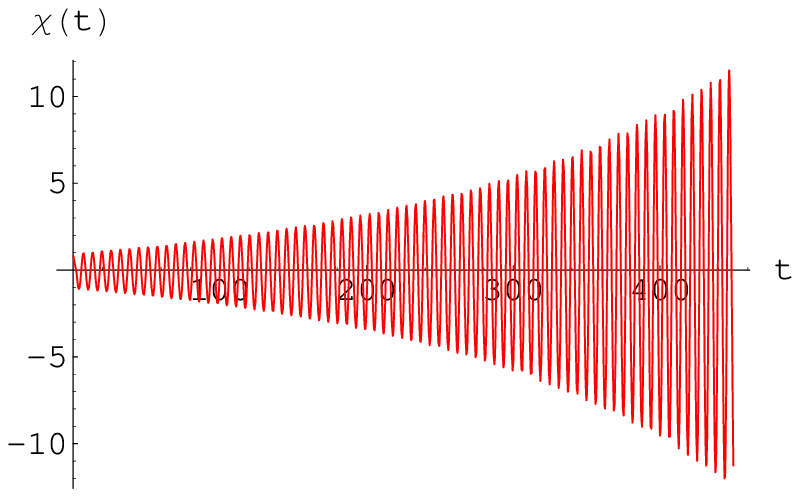} &  &
\includegraphics[scale=0.8]{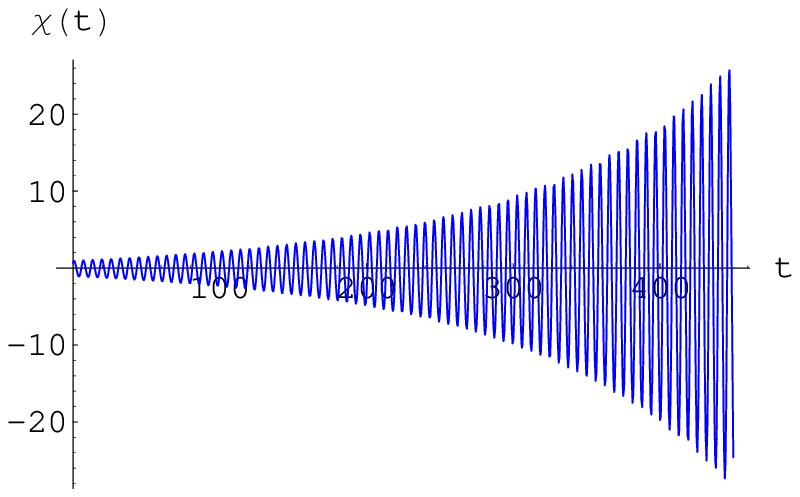}
\end{tabular}
\caption{\label{fig:chi} Oscillations of the field $\chi(t)$. {\it The left panel}
corresponds to harmonic potential of $\phi$ and {\it the right panel} corresponds to
the potential~\eqref{U-of-phi}. The time $t$ is measured in units of $m^{-1}$}
\end{figure}

Now we can numerically solve Eq.~\eqref{ddot-chi-flat} with the computed $\phi(t)$
and with the chosen values for this example: $m_\chi =0$, $k=1$, $g=5\cdot 10^{-8}$
(in units of $m$), and the initial conditions $\chi(0)=1/\sqrt{2\omega(0)}\approx
0.67$, $\dot\chi(0)=\sqrt{\omega(0)/2}\approx 0.745$, where $\omega^2(0)=k^2+g
\phi(0)$. These initial conditions correspond to the solution of
Eq.~\eqref{ddot-chi-flat} with constant $\omega$: $\chi(t)=e^{-i\omega
t}/\sqrt{2\omega V}$, where $V=1$ (in units of $m^{-3}$) is the volume. As a result
we obtain  function $\chi(t)$ shown in Fig.~\ref{fig:chi}. As expected, the
amplitude of oscillations of $\chi$ increases faster in the case of the
potential~\eqref{U-of-phi}; e.g. at $t=450\,m^{-1}$ the amplitude ratio is
approximately 2.

Since we are going to apply the results for the calculation of the particle
production rate at the end of cosmological inflation, it would be appropriate to
present the number density of the produced $\chi$-particles, $n_k(t)$, which is
defined as
\begin{equation}\label{n(t)}
n_k(t)=\left(\frac{\dot\chi^2}{2} +
\frac{\omega^2_k\chi^2}{2}\right)\frac{1}{\omega_k}\,,
\end{equation}
where the expression in the brackets is the energy of the mode with momentum $k$,
and $\omega_k=\sqrt{k^2+g \phi}$ is the energy of one $\chi$-particle.

\begin{figure}[h]
\center
\includegraphics[scale=0.8]{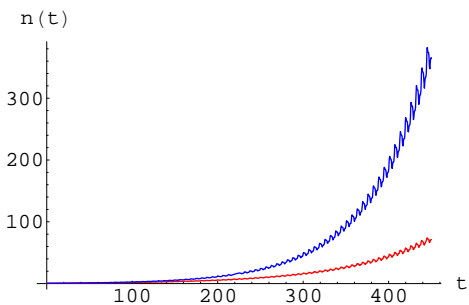}
\caption{\label{fig:n} The number density of produced $\chi$-particles $n_k(t)$. The
time $t$ is measured in units of $m^{-1}$}
\end{figure}

The number densities of the produced $\chi$-particles for the harmonic $\phi$ and
slightly step-like one are presented in Fig.~\ref{fig:n} by red and blue curves,
respectively.

Despite a decrease of particle number densities in some short time intervals, there
is an overall exponential rise, which goes roughly as $\chi^2$. Therefore, the ratio
of particle numbers for the two types of the potential is approximately equal to the
ratio of the amplitude squared, which is about 4 at $t=450\,m^{-1}$.

\section{Resonance in expanding universe}\label{s-res-exp}

The universe's heating after inflation was achieved  due to coupling of the inflaton
field $\phi$ to elementary particle fields. In this process mostly particles  with
masses smaller than the frequency of the inflaton oscillations were produced. The
decay of the inflaton could create both bosons and fermions. The boson production
might be strongly enhanced due to excitation of the parametric resonance in the
production process~\cite{ad-dk, brand-1, brand-2, kls-1, kls-2}. Hence bosons were
predominantly created initially. Later in the course of thermalization they gave
birth to fermions. For a model description of the first stage of this process we
assume, as we have done in Sect.~\ref{s-res-stat}, that the inflaton coupling to a
scalar field $\chi$ has the form $-g \phi \chi^2/2$, where $g>0$ is a coupling
constant with the dimension of mass. We also assume that the initial value of $\phi$
is positive. With this choice of the parameters the tachyonic situation can be
avoided. Otherwise $\chi$ would explosively  rise even at inflationary stage.  As a
result the contribution of $\chi$ to the total cosmological energy density would
become non-negligible and should be taken into account in the Hubble parameter. This
effect may inhibit inflation. These problems will be studied elsewhere. Below we
study a simpler situation of the initial stage of heating when the energy density of
the produced particles is small in comparison with the energy density of the
inflaton, so the back reaction is not of much importance. The effects of the back
reaction of particle production on the inflaton evolution are discussed in
Sect.~\ref{s-back}.

The equation of motion of the Fourier mode of $\chi$ with conformal momentum $k$ in
the FLRW metric has the form
\begin{equation}\label{ddot-chi}
\ddot \chi + 3 H \dot \chi + \left(\frac{k^2}{a^2} + g \phi\right) \chi = 0,
\end{equation}
where $a=a(t)$ is the cosmological scale factor, $H=\dot a/a$ is the Hubble
parameter, and field $\chi$ is taken for simplicity to be massless, $m_\chi=0$. We
assume that the universe is 3D-flat and that the cosmological energy density is
dominated by the inflaton field, so $H$ is expressed through $\phi$ as
\begin{equation}\label{H-of-phi}
H = \sqrt{\frac{8\pi}{3}}\, \frac{\sqrt{\dot\phi^{\,2}/2 + U(\phi)} }{m_{Pl}},
\end{equation}
where $U(\phi)$ is the potential of the inflaton, $m_{Pl} \approx 1.2 \cdot 10^{19}$
GeV is the Planck mass, and it is assumed, as usually, that the inflaton field is
homogeneous, $\phi = \phi (t)$. Correspondingly the equation of motion for $\phi$
has the form
\begin{equation}\label{eq:ddot-phi}
\ddot \phi + 3 H\dot \phi + U'(\phi) = 0,
\end{equation}
where $U' =dU/d\phi$.

We study here the particle production by $\phi$, which evolves in the
potential~\eqref{U-of-phi}, described in the previous section, where it has been
shown that the particle production in flat space-time is strongly enhanced in
comparison with the particle production by $\phi$ with the potential $U(\phi) = m^2
\phi^2/2$. We do the same thing here.

In expanding universe the liquid friction term $3 H \dot \phi$ in the equation of
motion for $\phi(t)$~\eqref{eq:ddot-phi} strongly modifies the evolution of $\phi$
and it is necessary that the resonance should be generated faster than $\phi$
significantly dropped down.

\begin{figure}[h]
\center
\begin{tabular}{c c c}
\includegraphics[scale=0.8]{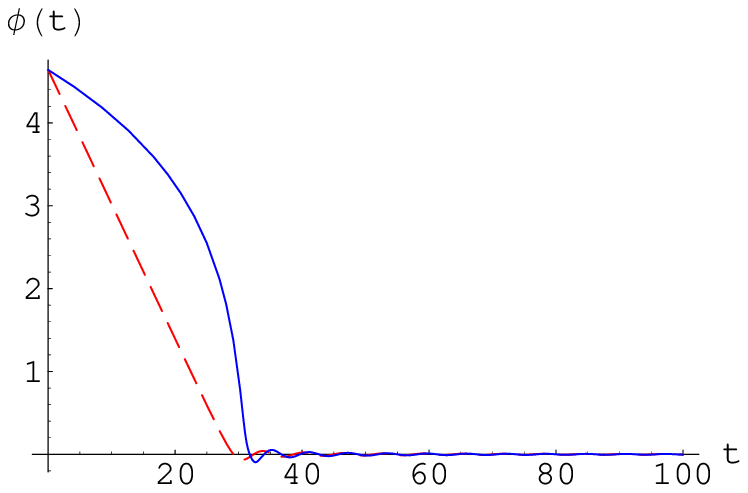} &  &
\includegraphics[scale=0.8]{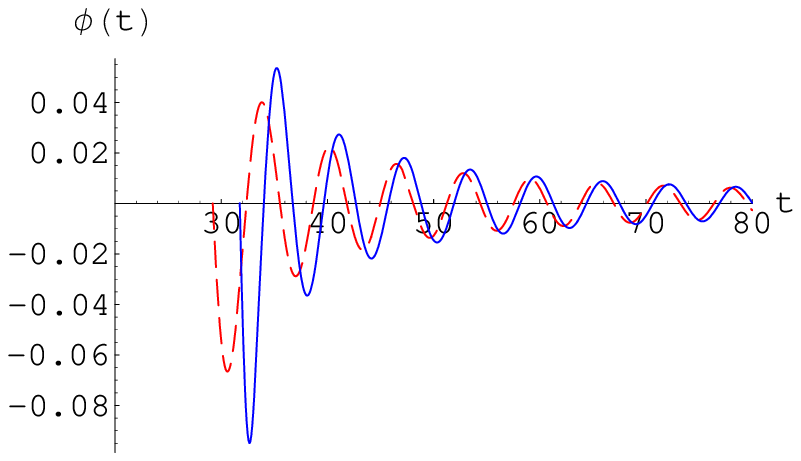}\\
\includegraphics[scale=0.8]{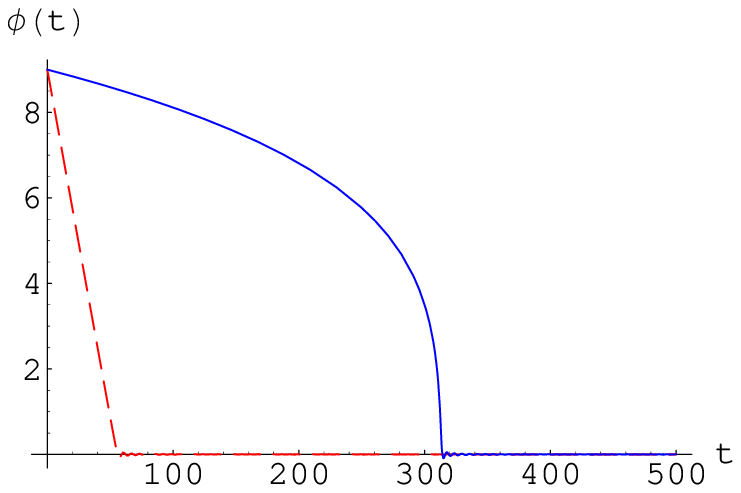} &  &
\includegraphics[scale=0.8]{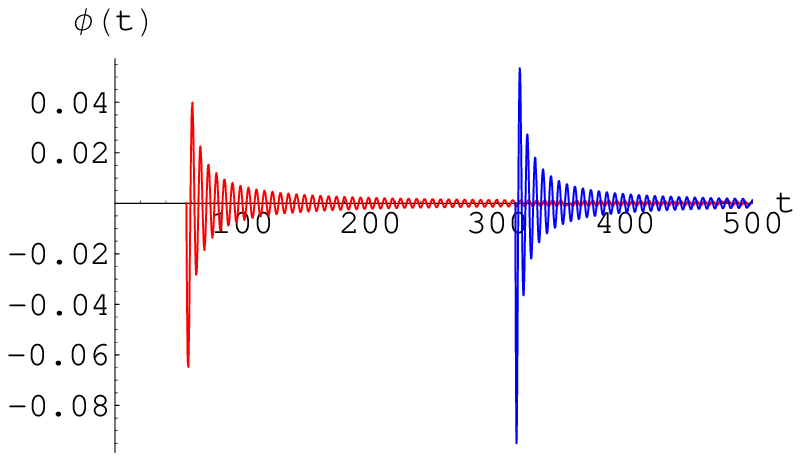}
\end{tabular}
\caption{\label{fig:phiH} Inflaton field, $\phi(t)$, in the expanding universe. The
time $t$ is measured in units of $m^{-1}$ and field $\phi$ is measured in units of
$m_{Pl}$. In the {\it left panels} the evolution $\phi(t)$ is shown starting from
$t=0$. In the {\it right panels} the oscillations of $\phi(t)$ are presented in more
detail, starting from the moment when $\phi = 0$ for the first time. {\it The upper
plots} correspond to the case $\phi(0)= 4.64\,m_{Pl}$, and {\it the lower ones}
correspond to the case $\phi(0)= 9\,m_{Pl}$}
\end{figure}

We find numerical solutions of Eq.~\eqref{eq:ddot-phi} with the same potentials as
above, i.e the harmonic one and $U(\phi)$ presented in Eq.~\eqref{U-of-phi}. As
initial conditions we take $\dot\phi(0)=0$ and two different values  $\phi(0)=
4.64\,m_{Pl}$ and $\phi(0)= 9\,m_{Pl}$, which we consider in parallel. The first
one, $\phi(0)= 4.64\,m_{Pl}$, is equal to the value which we took in the flat
universe case (see Fig.~\ref{fig:phi}). However, in this case the initial energies,
$U(\phi(0))$, are very different for the two potentials (see Fig.~\ref{fig:U}). So
we consider also the initial condition $\phi(0)= 9\,m_{Pl}$ for which the two
potentials $U(\phi(0))$ have equal magnitudes. The results of a numerical solution
of Eq.~\eqref{eq:ddot-phi} are shown in Fig.~\ref{fig:phiH}.

\begin{figure}[h]
\center
\begin{tabular}{c c c}
\includegraphics[scale=0.8]{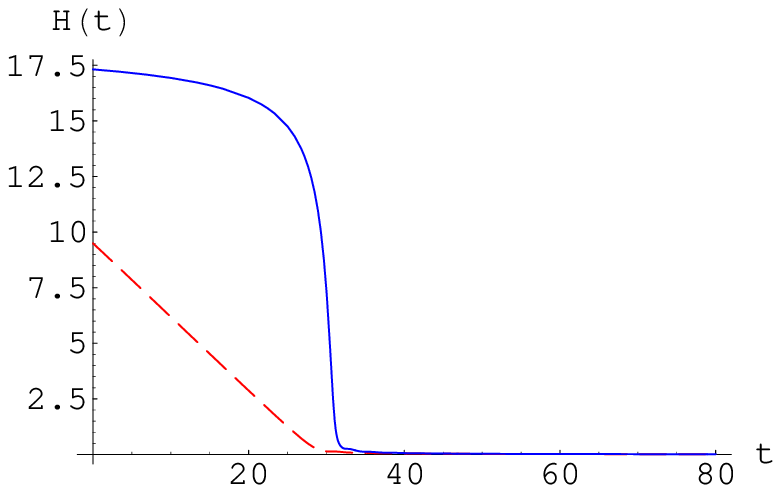} &  &
\includegraphics[scale=0.8]{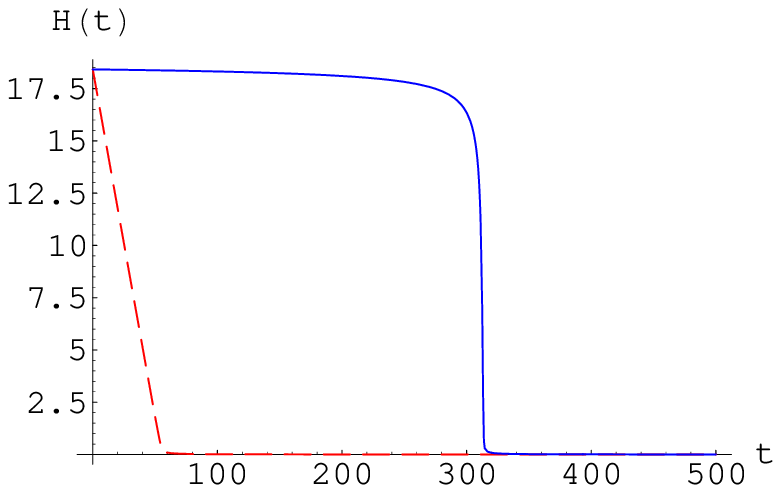}
\end{tabular}
\caption{\label{fig:H} Hubble parameter $H(t)$. The time $t$ is measured in units of
$m^{-1}$ and $H$ is measured in units of $m$. {\it The left plot} corresponds to the
case $\phi(0)= 4.64\,m_{Pl}$, and {\it the right one} corresponds to the case
$\phi(0)= 9\,m_{Pl}$}
\end{figure}

\begin{figure}[h]
\center
\begin{tabular}{c c c}
\includegraphics[scale=0.8]{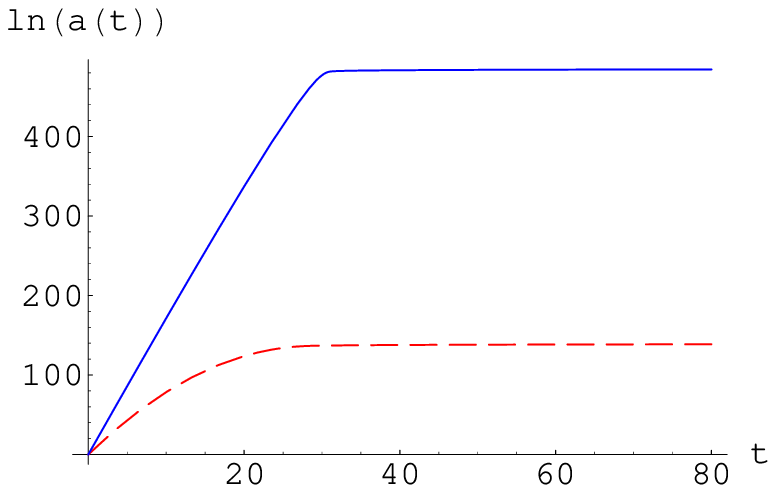} &  &
\includegraphics[scale=0.8]{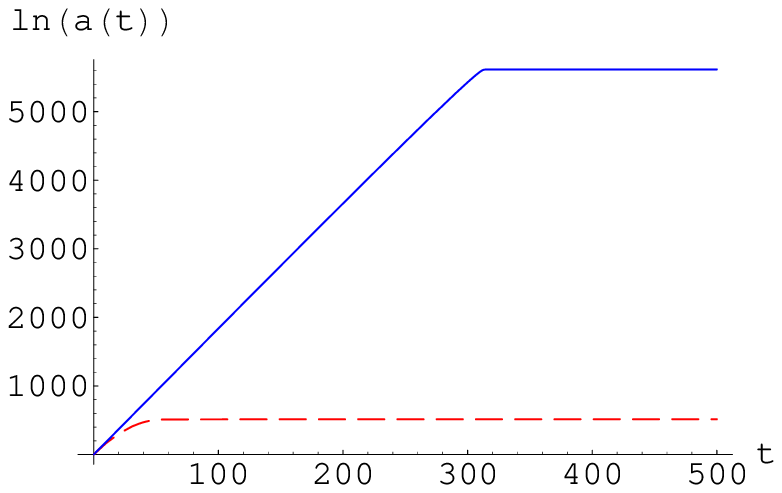}
\end{tabular}
\caption{\label{fig:a} Logarithm of the scale factor $a(t)$. The time $t$ is
measured in units of $m^{-1}$. The initial value is $a(0)=1$. {\it The left plot}
corresponds to $\phi(0)= 4.64\,m_{Pl}$, and {\it the right one} corresponds to
$\phi(0)= 9\,m_{Pl}$}
\end{figure}

At the stage of inflation the inflaton field, $\phi(t)$, rolls toward the minimum of
potential quite slowly, due to the large "friction" $H$. For successful inflation
one needs the condition $\int_{t_i}^{t_e} H(t)dt > 70$ to be satisfied, where $t_i$
is the time of the beginning and $t_e$ is the time of the end of inflation. It can
easily be seen in Fig.~\ref{fig:H} that this condition is fulfilled for both
potentials. The scale factor $a(t)$ grows exponentially during inflation (see
Fig.~\ref{fig:a}).

After  $\phi(t)$ reaches the minimum of the potential, it does not have enough
energy to climb high back because of the energy loss due to the friction, so $\phi$
starts to oscillate with decreasing amplitude. The moment $t_0$, when $\phi(t)$
crosses zero for the first time, can be considered as the end of inflation, the
onset of the oscillations, and the universe's heating. The Hubble parameter becomes
quite small by this moment, $H \lsim m$, and continues to decrease, so during the
oscillation period one can neglect $H$ in comparison with $m$.

Equation~\eqref{eq:ddot-phi} is simplified by the substitution
$\phi(t)=\Phi(t)/a^{3/2}(t)$, and in the case of quadratic potential it has a
solution $\phi(t)=\phi_0(t)\cos mt$, where $\phi_0(t)\sim a^{-3/2}(t)$. Therefore,
Eq.~\eqref{ddot-chi} turns into
\begin{equation}\label{ddot-chi-1}
\ddot \chi + 3H \dot\chi + \frac{k^2}{a^2}\left(1+\frac{g\phi_0 a^2}{k^2}\cos
mt\right) \chi = 0,
\end{equation}
which is the Mathieu equation with  a friction term, so the condition of parametric
resonance~\cite{ll} is
\begin{equation}\label{condition}
\left|m-\frac{2k}{a}\right|<\sqrt{\left(\frac{g\phi_0 a}{2k}\right)^2-9H^2}.
\end{equation}

Let us consider now Eq.~\eqref{ddot-chi} in the general case. It is convenient to
make the substitution $\chi(t)=X(t)/a^{3/2}(t)$, so one obtains
\begin{equation}\label{ddot-X}
\ddot X + \left(\frac{k^2}{a^2} + g\phi-\frac{3}{4}H^2-\frac{3\ddot a}{2a}\right) X
= 0.
\end{equation}

Equation~\eqref{ddot-X} has the form of the free oscillator equation $\ddot X +
\omega^2 X=0$ with frequency depending on time. During the oscillations the terms
$H^2$ and $\ddot a/a$ are relatively small, so one can take
\begin{equation}\label{omega_chi}
\omega(t) \approx \sqrt{\frac{k^2}{a^2(t)} + g\phi(t)}.
\end{equation}
If we neglect the time dependence of $a(t)$ and take a small $g$, then $\omega$
would be almost constant and the solution of~\eqref{ddot-X} is $X(t)\simeq
e^{-i\omega t}/\sqrt{2\omega}$, which corresponds to $\chi(t)=e^{-i\omega
t}/\sqrt{2\omega V}$, where $V=a^3$ is the comoving volume.

The energy and number of produced particles in a comoving volume are, respectively,
\begin{equation}\label{E_X}
E(t)=\frac{\dot X^2}{2} + \frac{\omega^2 X^2}{2},
\end{equation}
\begin{equation}\label{N_X}
N(t)=\left(\frac{\dot X^2}{2} + \frac{\omega^2 X^2}{2}\right) \frac{1}{\omega}.
\end{equation}

The scale factor $a(t)$ changes much slower with time than $X(t)$, therefore
$\dot\chi\approx \dot X/a^{3/2}$. Thus the energy and the number densities of the
produced $\chi$-particles would be
\begin{equation}\label{rho}
\rho_\chi(t)=\frac{\dot \chi^2}{2} + \frac{\omega^2 \chi^2}{2},
\end{equation}
\begin{equation}\label{n}
n_\chi(t)=\left(\frac{\dot \chi^2}{2} + \frac{\omega^2
\chi^2}{2}\right)\frac{1}{\omega}.
\end{equation}

It is reasonable to impose the initial conditions for the field $\chi(t)$ at the
moment $t_{0}$, which is the moment of the onset of the inflaton oscillations. We
choose the initial conditions as $\chi(t_{0})=1/\sqrt{2\omega(t_{0})V(t_{0})}$,
$\dot \chi(t_{0})=\sqrt{\omega(t_{0})/2V(t_{0})}$, where $V(t_{0})=1$ in units of
$m^{-3}$. These conditions correspond to vanishing initial density of the
$\chi$-particles. To avoid the tachyonic situation we put $k/a(t_{0})=50\,m$ and
$g=5\cdot10^{-4}\,m$, therefore we ensure that $\omega^2$ is always positive during
the interesting time interval.

\begin{figure}[h]
\center
\begin{tabular}{c c c}
\includegraphics[scale=0.8]{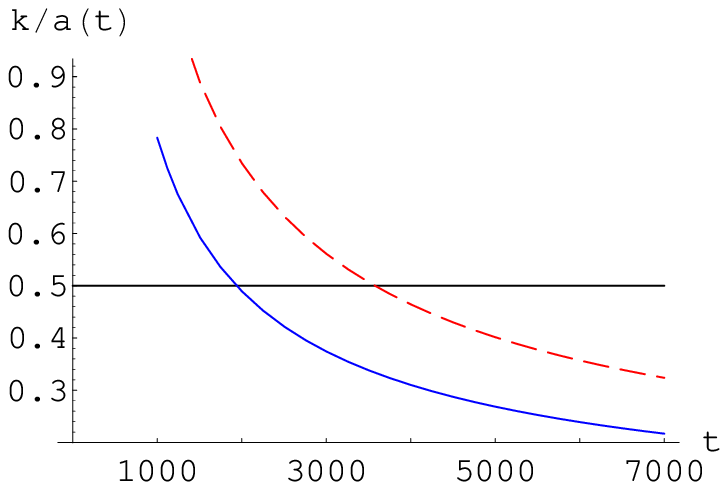} &  &
\includegraphics[scale=0.8]{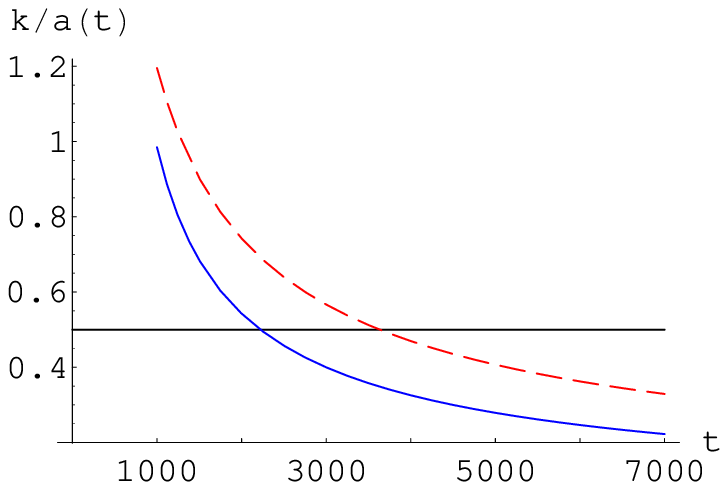}
\end{tabular}
\caption{\label{fig:k} The physical momentum $k/a(t)$ (in units of $m$). {\it The
black horizontal lines} cross the functions $k/a(t)$ in the resonance points,
$k/a(t)=m/2$. {\it The left plot} corresponds to $\phi(0)= 4.64\,m_{Pl}$, and {\it
the right one} corresponds to $\phi(0)= 9\,m_{Pl}$}
\end{figure}

At large time the amplitude of the $\phi$ oscillation becomes small and the
potential~\eqref{U-of-phi} closely approaches the quadratic one; therefore the
condition of parametric resonance~\eqref{condition} holds for both potentials. Thus,
the resonance occurs in the narrow region when $k/a(t)$ is near $m/2$. In
Fig.~\ref{fig:k} one can see at what time the resonance occurs for the two
potentials of $\phi$ with the chosen parameters.

We have assumed here that the inflaton field gives the dominant contribution to the
cosmological energy density. Therefore, when the energy density of the produced
particles, $\rho_\chi$, becomes comparable to $\rho_\phi=\dot \phi^2/2+U(\phi)$, the
model stops to be self-consistent and we should modify the calculations. The easiest
way is to take the energy density of the produced particles at this moment as an
ultimate one and to estimate the cosmological heating temperature on the basis of
this result. A more precise way is to take into account the back reaction of the
produced particles on the damping of the inflaton oscillations and to include the
contribution of the created particles into the Hubble parameter. The first
simplified approach, which gives a correct order of magnitude estimate of the
temperature,  is sufficient for our purposes.

Resonance particle production could induce specific features in the primordial
spectrum of density perturbations~\cite{delta-rho-res}, which might be potentially
observable. We thank the referee for mentioning this effect. This could be the
subject of a separate study.

\begin{figure}[h]
\center
\begin{tabular}{c c c}
\includegraphics[scale=0.8]{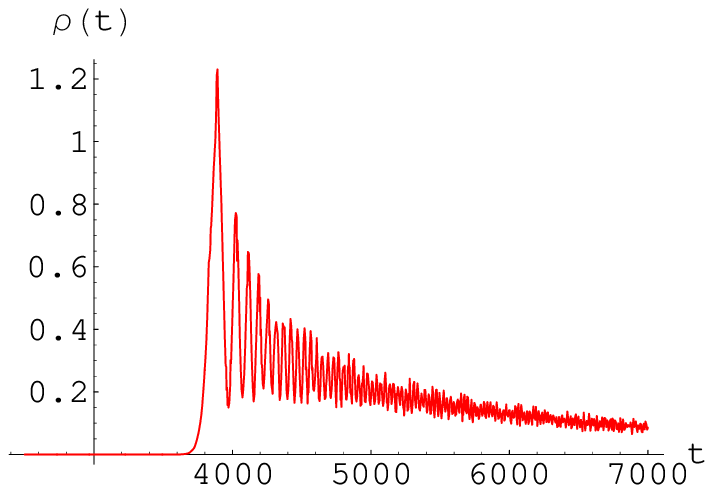} &  &
\includegraphics[scale=0.8]{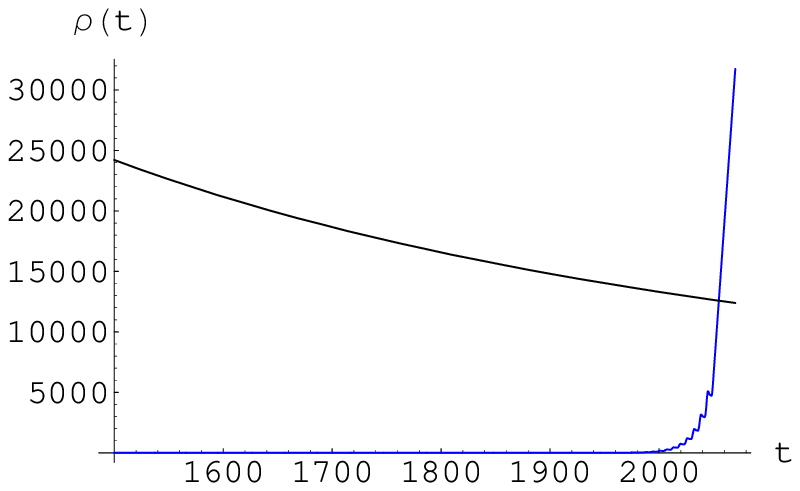}\\
\includegraphics[scale=0.8]{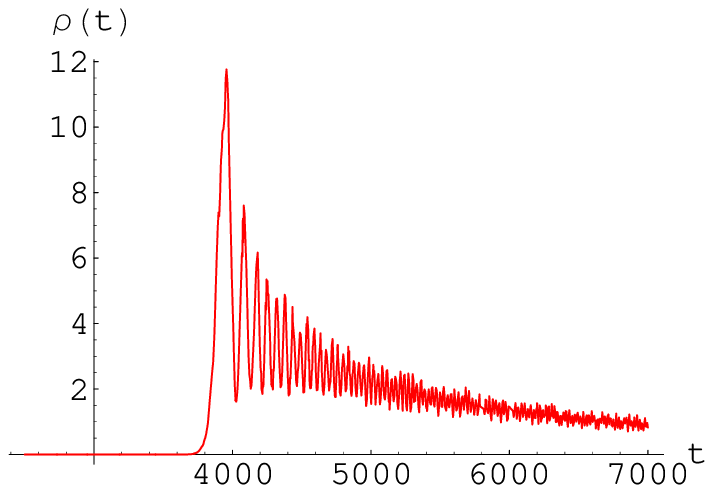} &  &
\includegraphics[scale=0.8]{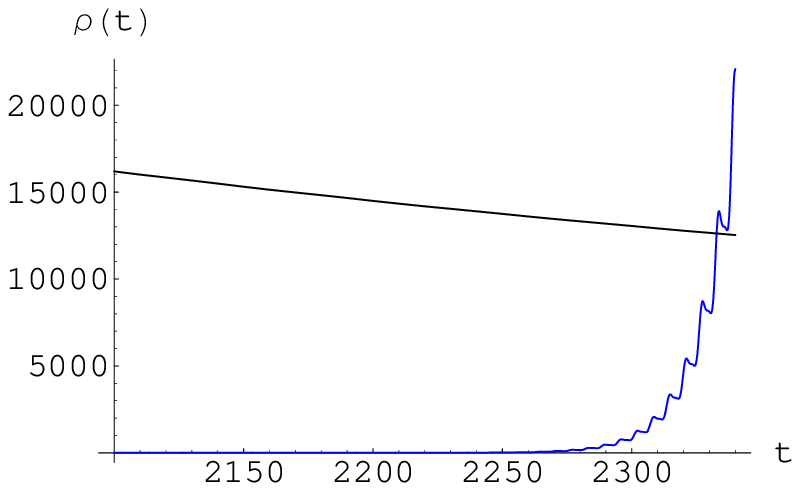}
\end{tabular}
\caption{\label{fig:rho} Energy density, $\rho(t)$, of the produced particles
$\chi$. The time $t$ is measured in units of $m^{-1}$ and $\rho(t)$ is measured in
units of $m^4$. {\it The upper plots} correspond to $\phi(0)= 4.64\,m_{Pl}$, and
{\it the lower ones} correspond to $\phi(0)= 9\,m_{Pl}$. {\it Black lines in the
right panels} denote $\rho_\phi$}
\end{figure}

The energy densities of the produced particles, $\rho(t)$, for
$\phi(0)=4.64\,m_{Pl}$ and $\phi(0)= 9\,m_{Pl}$ are presented in Fig.~\ref{fig:rho}.
One can see that for the quadratic potential of $\phi$ the parametric resonance is
quite weak and the energy density of particles produced during the resonance is much
less than $\rho_\phi$, which is about $3\cdot 10^3\,m^4$ at the moment of the
resonance emergence. On the contrary, in the potential~\eqref{U-of-phi} $\rho_\chi$
increases very quickly and becomes comparable to  $\rho_\phi$ at  $t \approx 2050$
and $t \approx 2330$ in the cases $\phi(0)= 4.64\,m_{Pl}$ and $\phi(0)= 9\,m_{Pl}$,
respectively.

\begin{figure}[h]
\center
\begin{tabular}{c c c}
\includegraphics[scale=0.8]{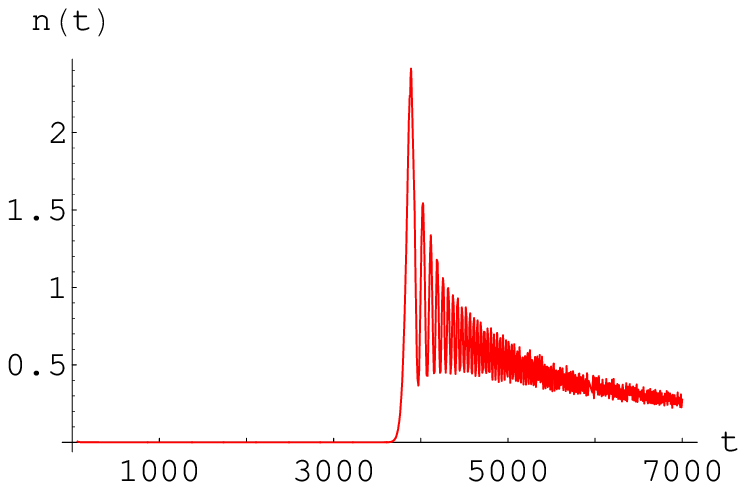} &  &
\includegraphics[scale=0.8]{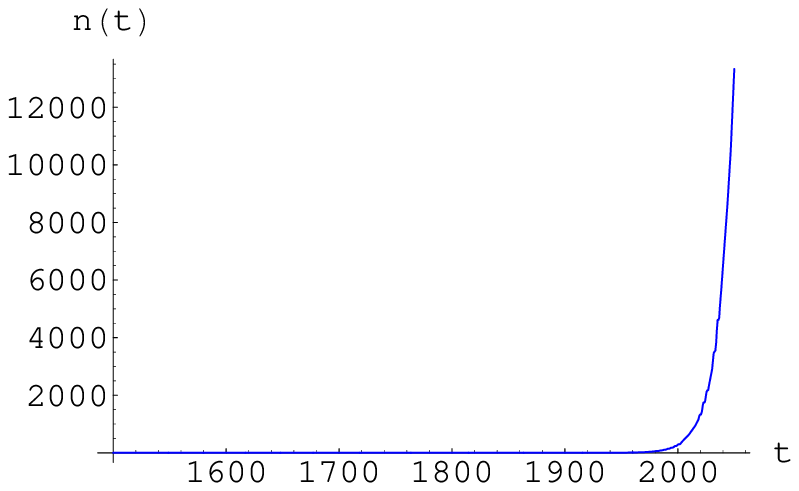}\\
\includegraphics[scale=0.8]{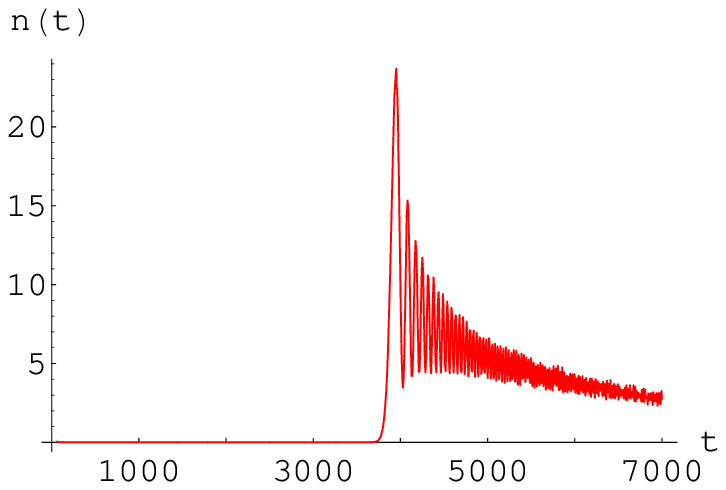} &  &
\includegraphics[scale=0.8]{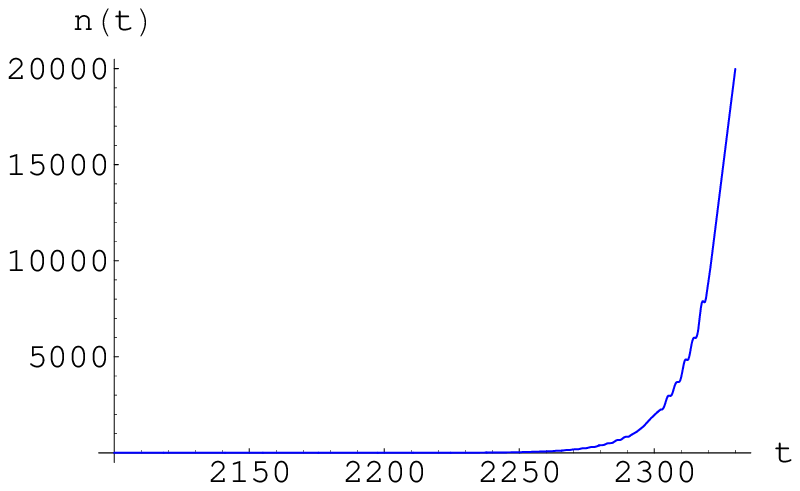}
\end{tabular}
\caption{\label{fig:nH} The number density of the produced $\chi$-particles $n(t)$.
The time $t$ is measured in units of $m^{-1}$ and number density is measured in
units of $m^3$. {\it The upper plots} correspond to $\phi(0)= 4.64\,m_{Pl}$ and {\it
the lower ones} correspond to $\phi(0)= 9\,m_{Pl}$}
\end{figure}

\begin{figure}[h]
\center
\begin{tabular}{c c c}
\includegraphics[scale=0.8]{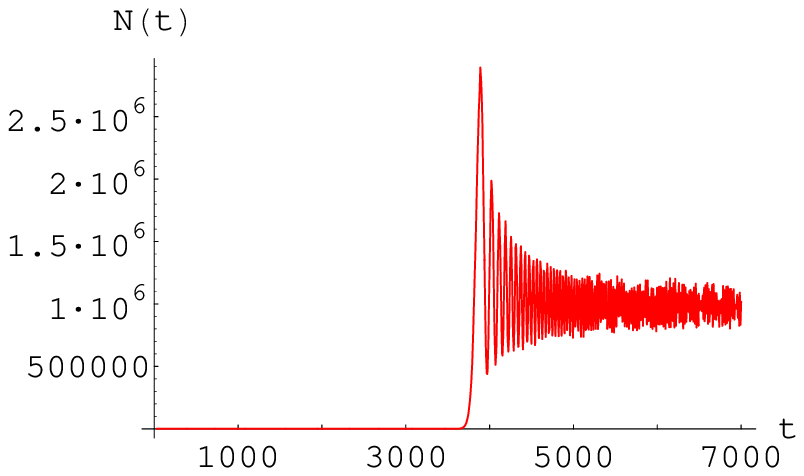} &  &
\includegraphics[scale=0.8]{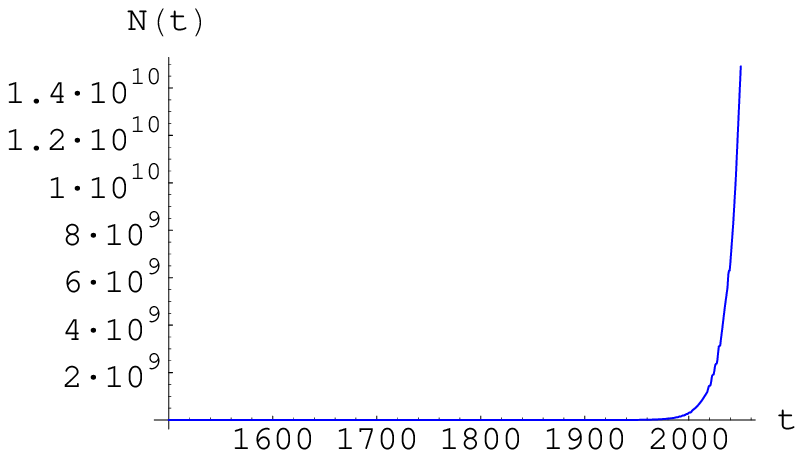}\\
\includegraphics[scale=0.8]{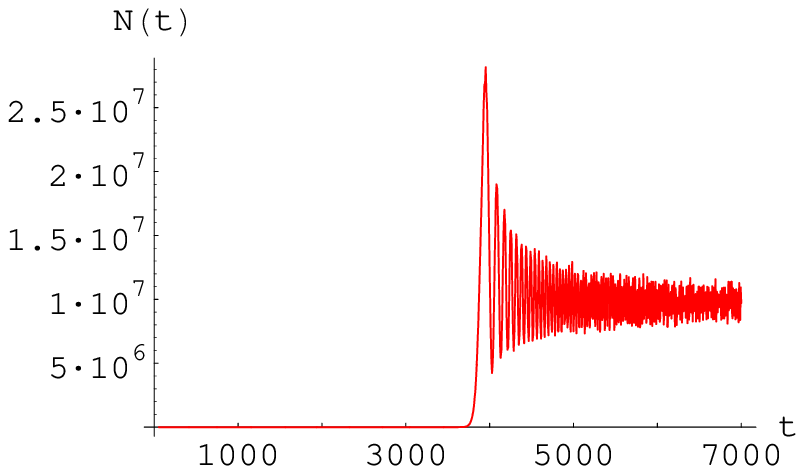} &  &
\includegraphics[scale=0.8]{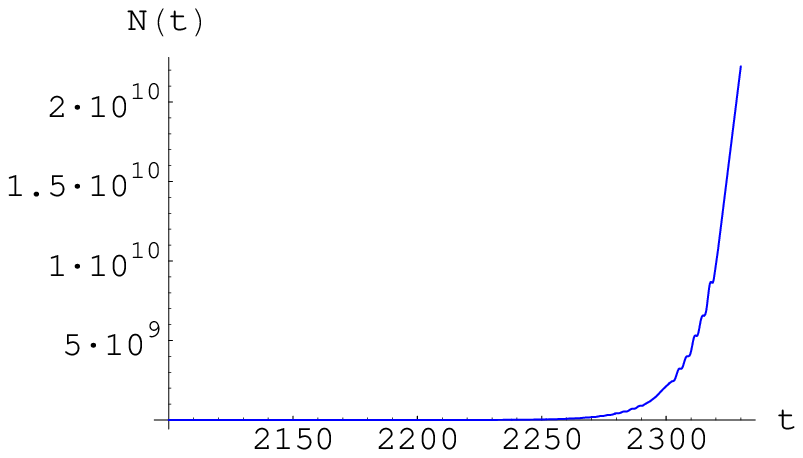}
\end{tabular}
\caption{\label{fig:N} The total number of produced $\chi$-particles $N(t)$. The
time $t$ is measured in units of $m^{-1}$. {\it The upper plots} correspond to
$\phi(0)= 4.64\,m_{Pl}$, and {\it the lower ones} correspond to  $\phi(0)=
9\,m_{Pl}$}
\end{figure}

The number densities $n(t)$ calculated according to Eq.~\eqref{n} and the total
numbers $N(t)=n(t)\cdot a^3(t)$ of the produced particles are presented in
Figs.~\ref{fig:nH} and \ref{fig:N}, respectively. These plots can be understood as
follows. During some time after the beginning of the $\phi$ oscillations the
conditions of parametric resonance are not fulfilled and therefore $\chi$-particles
are not produced in a considerable amount. Due to the universe's expansion the term
$(k/a(t))^2$ drops down and at some moment the mode with conformal momentum $k$
enters the resonance region. It is  manifested as an exponentially growth of
$\rho(t)$, $n(t)$, and $N(t)$. Then the resonance conditions stop to be satisfied
once again and $\chi$-particle production is almost terminated, so their total
number tends to a constant value, while their number density, $n(t)$, decreases
as~$a^{-3}$.

Let us suppose that $\chi$-particles reach thermal equilibrium very quickly. The
energy density of relativistic particles is simply related to the temperature:
\begin{equation}
\rho_\chi=\frac{\pi^2}{30}g_* T^4,
\end{equation}
where $g_*\sim 100 $ is the number of relativistic particle species in the
thermalized plasma. Therefore, from Fig.~\ref{fig:rho} in the case of the
potential~\eqref{U-of-phi} one can estimate the temperature by the moment when
$\rho_\chi \sim \rho_\phi$ as
\begin{equation}
T \sim 5 \,m \approx 10^{-5}\,m_{Pl} \sim 10^{14}\, \rm{GeV}.
\end{equation}

It is instructive to present a qualitative explanation of the obtained results. The
initial moment, when the particle production is initiated, is the moment of the
onset of the inflaton oscillations. They started when the Hubble friction term in
the inflaton equation of motion (\ref{eq:ddot-phi}) becomes small in comparison with
the potential term. It took place when $H^2$ dropped down below $U''(\phi)$. For the
harmonic potential $U_h = m^2 \phi^2 /2$ it happened after $\phi$ reached the
boundary value $\phi_h^2 = {3m_{Pl}^2}/{4\pi}$. For our potential (\ref{U-of-phi})
the oscillation regime is reached at a $\sqrt{6}$ times larger value of $\phi$, i.e.
$\phi^2_\lambda = 9m_{Pl}^2/2\pi$. This result is obtained in the case of $\lambda_0
\gg 1$, as has been chosen for the potential~(\ref{U-of-phi}).

Discussing particle production we considered the initial physical momentum of the
produced particles $p=k/a_{in} = 50\,m$ to avoid a tachyonic excitation, and we took
initial conditions for $\chi$ corresponding to the vacuum state, i.e. to the state
where the real $\chi$-quanta were absent. In the course of the cosmological
expansion this momentum evolved down to the resonance value $k/a_{res} = m/2$, both
in the cases of harmonic and non-harmonic evolution. Since the Hubble parameter in
the modified theory was larger than that in the harmonic case, the former reached
the resonance value faster; see Fig.~\ref{fig:k}. However, this is not directly
essential, because in both cases $p$ reaches the resonance value at the same
redshift relative to the initial state. An important factor which determines the
effectivity of the resonance particle production is the amplitude of the inflaton
field at the resonance. Initial amplitudes differed by the factor $\sqrt{6}$, but
this is not the end of the story. The amplitude of the "harmonic inflaton" dropped
down as $1/a^{3/2}$, while the evolution of the inflaton living in the potential
(\ref{U-of-phi}) is considerably slower. For a purely quartic potential, $\sim
\lambda \phi^4$, the amplitude of the inflaton field drops as $1/a$, so it comes to
the resonance with the amplitude 10 times larger than the harmonic inflaton.

In the case considered in this paper the modified potential is not purely quartic,
moreover, it approaches the harmonic form at small $\phi$, when the resonance is
excited. Nevertheless, the amplitude of $\phi$ during the harmonic regime is much
larger than in the purely harmonic case, though not by such a large factor.

According to the equations presented in Sect.~\ref{s-res-stat} the amplitude of
$\chi$ at the resonance rises as $\sim \exp (g \phi_{res} t/2m)$, where $\phi_{res}$
is the inflaton amplitude at the resonance. This explains the difference in
efficiency of the particle production  between the purely harmonic and modified
cases.

Finally, it should be noted that production of $\chi$-particles is possible also due
to the usual non-resonant decay $\phi \rightarrow \chi\chi$. However, the width
$\Gamma$ of such a decay is very small. Indeed, it follows from Eq.~\eqref{L-int}
that $\Gamma=g^2m/32\pi$ for neutral massless $\chi$. Therefore, for our choice of
the interaction constant, $g=5\cdot 10^{-4}\,m$, the typical time of the decay is
$\tau=1/\Gamma \sim 4\cdot 10^8\,m^{-1}$, which is much longer than the time when
the resonance occurs (see Fig.~\ref{fig:k}). Thus the contribution of the
non-resonant decay is negligible.

The inflationary model based on the potential~\eqref{U-of-phi} is similar to the
well-known new inflationary scenario or inflation at a small field $\phi$; see e.g.
Ref.~\cite{Gorbunov}. Correspondingly, the slow roll parameters satisfy the
condition $\varepsilon \ll \left|\eta\right|$. The definition of the parameters and
their relation to scalar and tensor perturbations can be found in
Refs.~\cite{Gorbunov, LL, pdg, baumann}. The inequality mentioned above means that
the amplitude of the gravitational waves in the model considered here must be very
small. However, this conclusion is model dependent and is not necessarily true.

\section{Back reaction \label{s-back}}

The tremendous rate of particle production makes its back reaction on the inflaton
evolution non-negligible quite soon. The simplest way to estimate this back reaction
is to use the density balance condition: the energy loss by the inflaton must be
equal to the energy carried away by the produced particles. Such a comparison is
done in Fig.~\ref{fig:rho}. Before this moment we can neglect the related decrease
of the inflaton amplitude. This is similar to the well-known instant decay
approximation, which usually works pretty well. So as regards the order of magnitude
we can rely on the results obtained with the neglected back reaction. However, in
the instant decay approximation the decay rate remains constant, while the
parametric resonance rate is proportional to the amplitude of the inflaton, and so
with decreasing $\phi$ the resonance production drops down and it may happen that
the remaining part of the inflaton energy would decay much more slowly. Keeping in
mind that usually the inflaton makes nonrelativistic matter, while produced
particles are relativistic, we can conclude that the relative contribution of the
inflaton to the cosmological energy density would grow and ultimately the dominant
part of the (re)heated cosmic plasma would be created by slow (perturbative)
inflaton decay. However, this is not a subject of the present work.

For a more accurate treatment of the back reaction we can use the equation of motion
of the inflaton with quantum effects induced by particle production in the one loop
approximation~\cite{ad-sh, ad-kf}. The corresponding equation is an
integro-differential one, which for the coupling of the produced quantum field
$\chi$ of the form (\ref{L-int}) can be written as \be &&\ddot{\vpc} + V'(\vpc) -
 \frac{g^2}{16 \pi^2}\,\vpc \ln { t_1\over\epsilon } \nonumber \\
&&= \frac{g^2}{16 \pi^2} \int_0^{t_1} \frac{d \tau}{\tau} \, \left[ \vpc(t-\tau) -
\vpc(t) \right] + \frac{g^2}{16 \pi^2} \int_{t_1}^{t-t_{in}} \frac{d \tau}{\tau} \,
 \vpc(t-\tau)~.
\label{renchi} \ee Here we use the notation $\vpc$ for the classical inflaton field
to keep track with Ref.~\cite{ad-sh}. The integrals in the r.h.s. are ultraviolet
finite. The logarithmically infinite contribution in the l.h.s., related to the
ultraviolet cut-off $\epsilon \rightarrow 0$, is taken out by the mass
renormalization, so with a possible bare mass term in the potential, $V_{m_0} =
m_0^2 \vpc^2 /2$ (here $m_0$ is the bare mass of $\vpc$), we obtain \be &&
\ddot{\vpc} + m^2 (t_1) \vpc + \left[ V'(\vpc) -V'_m (\vpc) \right] \nonumber \\
&&= \frac{g^2}{16 \pi^2} \int_0^{t_1} \frac{d \tau}{\tau} \, \left[ \vpc(t-\tau) -
\vpc(t) \right] + \frac{g^2}{16 \pi^2} \int_{t_1}^{t-t_{in}} \frac{d \tau}{\tau} \,
 \vpc(t-\tau) ~,
\label{finchi} \ee where  $t_1$ is an arbitrary normalization point and the
"running" mass is $m^2 (t_1) = m^2 (t_2) - (g^2/16\pi^2) \ln (t_1/t_2)$. In
Eq.~(\ref{finchi}) we explicitly separated the massive part, $V_{m_0}$, in the
potential, so that the term in the square brackets vanishes for the harmonic
potential, $V(\vp) = m_0^2 \vp^2 /2$.

The equation governing the evolution of $\phi_c$ can be grossly simplified if the
particle production goes in the resonant mode. In this case the occupation numbers
of the field $\chi$ become very large and the field can be treated as a classical
one.

We assume, as is usually done in studies of  parametric resonance, that $\chi$ is a
real field. Since the resonance is rather narrow, the universe's expansion can be
neglected. Correspondingly, the $k$-mode of the massless field $\chi$ satisfies the
following equation of motion:
\begin{equation}
\ddot\chi_k(t) + k^2 \chi_k(t) + g \phi_c(t) \chi_k (t) = 0. \label{ddot-chi-k}
\end{equation}
In what follows we omit the subindices $k$ and $c$.

This equation can be transformed into the integral equation: \be \chi(t)  = \chi_0 -
g\,\int_0^t dt' G_R (t-t') \phi(t') \chi (t'), \label{int-eq} \ee where $\chi_0$ is
the initial value of $\chi$, which is assumed for simplicity to be zero (it does not
rise exponentially, so can be neglected anyhow) and \be G_R (t) = \left\{
\begin{array}{rl}
\sin k t /k, & \text{ if  }   t > 0,\\
0, & \text{ if } t < 0,
\end{array} \right.
\label{green-func} \ee is the retarded Green's function of Eq.~(\ref{ddot-chi-k}).

Making the ansatz $\chi=A\exp(\gamma t)\sin(kt+\alpha)$, and neglecting oscillating
terms in the integral (\ref{int-eq}), we find that this ansatz is self-consistent,
i.e. $\chi$ indeed rises exponentially, if $\phi = \phi_0 \sin mt$, and $k = m/2$,
$\alpha=\pi/2$:
\begin{equation}
\chi = A\,\frac{g\phi_0}{4k\gamma}\, e^{\gamma t} \cos kt. \label{chi-res}
\end{equation}
It is assumed here that both $A$ and the amplitude of the inflaton oscillations,
$\phi_0$, slowly change with time. For self-consistency one needs to impose the
condition $ g \phi_0 /4k \gamma = 1$. It leads to the canonical expression for
$\gamma$ presented in Eq.~(\ref{chi0-of-t}).

Let us turn now to Eq.~(\ref{ddot-phi-0}). The impact of the $\chi$-particle
production originating from the term $-g\chi^2/2$ can be estimated as follows. We
substitute expression (\ref{chi-res}) for $\chi$ and assume that $\phi$ evolves as
$\phi= \phi_0 (t) \sin mt $, where $\phi_0(t)$ is (in comparison with frequency $m$)
a slowly varying function of time. In this way we obtain the equation:
\begin{equation}
2m\dot\phi_0\cos mt = -\frac{1}{2}g A^2 e^{2\gamma t} \cos^2 kt = -\frac{gA^2}{4}
e^{2\gamma t} (1+\cos 2kt). \label{m-dot-phi}
\end{equation}
The first term in the brackets in the r.h.s. describes the tadpole contribution and
should be disregarded. The second term is consistent with the l.h.s. if $k=m/2$, as
expected, and ultimately we arrive at the equation
\begin{equation}
\dot\phi_0 = -\frac{gA^2}{8m} e^{2\gamma t}. \label{dot-phi}
\end{equation}
As $\dot\phi_0$ is always negative, the amplitude of the inflaton oscillation
$\phi_0$ constantly decreases. Thus, if we neglect variations of $\gamma$ and take
$\gamma=g\phi_{0,res}/2m$, where $\phi_{0,res}$ is the value of $\phi_0$ at the
beginning of resonance, we obtain the lower limit on the time of the oscillation
damping. Indeed, in such a case the amplitude of the inflaton oscillation can easily
be found:
\begin{equation}
\phi_0 = \phi_{0,res}-\frac{gA^2}{16m\gamma} e^{2\gamma t}, \label{phi_0}
\end{equation}
and the characteristic time of the oscillation damping is
\begin{equation}
\tau_d \approx \frac{1}{2\gamma} = \frac{m}{g \phi_{0,res}}. \label{tau_d}
\end{equation}
According to our calculations $\phi_{0,res} \approx  10^{-4}\,m_{Pl}=100\, m$, and
the parameter $A$ can be estimated as $A\sim 0.1\,m$ (the initial value of $\chi$).
Using also $g = 5 \cdot 10^{-4}\,m$, we find that $\tau_d \sim 20/m$.

\begin{figure}[h]
\center
\begin{tabular}{c c c}
\includegraphics[scale=0.8]{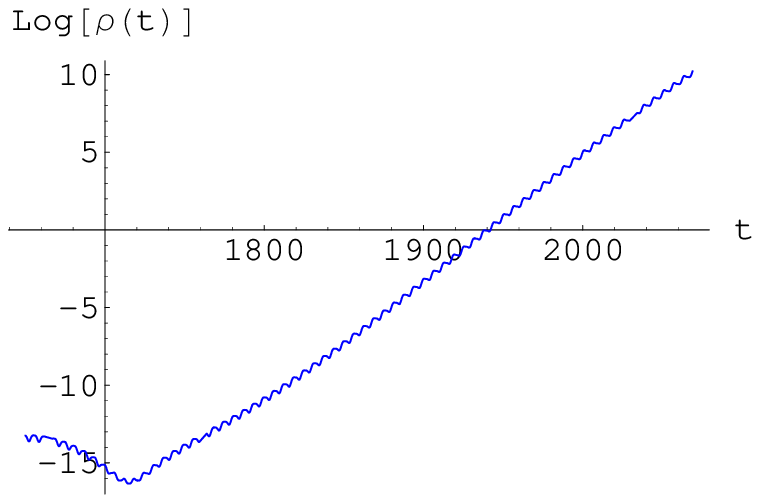} &  &
\includegraphics[scale=0.8]{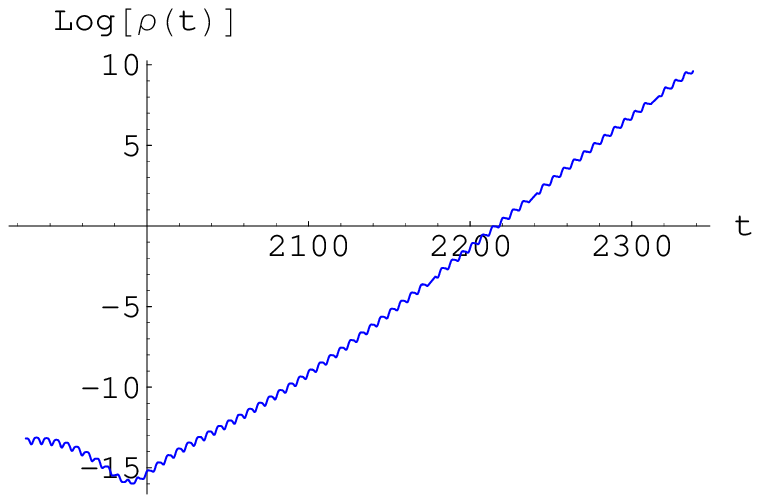}
\end{tabular}
\caption{\label{fig:logrho} Energy density of the produced $\chi$-particles for
$\phi(0)= 4.64\,m_{Pl}$ ({\it left panel}) and for $\phi(0)= 9\,m_{Pl}$ ({\it right
panel}) in a normal logarithm scale}
\end{figure}

In Fig.~\ref{fig:logrho} the energy density of the produced $\chi$-particles is
presented, as usually, for $\phi(0)= 4.64\,m_{Pl}$ (left panel) and for $\phi(0)=
9\,m_{Pl}$ (right panel). The exponential rise is clearly observed with the exponent
quite close to the above calculated one.

The characteristic lapse of time during which $\phi$ disappears down to zero can be
estimated from Eq.~(\ref{phi_0}) and is equal to
\begin{equation}
\tau_\phi \sim \frac{1}{2\gamma} \ln \frac{16m\gamma\phi_{0,res}}{gA^2} \sim
\frac{m}{ g \phi_{0,res}} \ln \frac{8\phi_{0,res}^2}{A^2}. \label{tau}
\end{equation}
With the chosen above parameter values we have $\tau_\phi \approx 300/m$. In reality
it would be somewhat longer because with the decreasing amplitude of $\phi$ the
parametric resonance exponent drops down proportionally to $\phi$.

\section{Conclusion} \label{s-conclud}

We have shown that even a small modification of the shape of the inflaton
oscillations could lead to a significant increase of the probability of particle
production by the inflaton and consequently to a higher universe temperature after
inflation. The particle (boson) production by the inflaton oscillating in a harmonic
potential was compared to the particle production in a toy inflationary model with a
flat inflaton potential at infinity. It was found that the parametric resonance in
the latter case is excited much more efficiently.

The inflationary model which is considered above is not necessarily realistic. We
took it as a simple example to demonstrate the efficiency of the particle production
and the heating of the universe. The impact of the energy density of the produced
particles on the cosmological expansion may noticeably change the phenomenological
properties of the underlying inflationary model.

We skip on purpose a possible tachyonic amplification of $\chi$-excitement to avoid
the deviation from the main path of the work on amplification of particle production
due to variation of the shape of the inflaton oscillations.

In Ref.~\cite{underwood} a similar study of the impact of the anharmonic corrections
to the inflaton oscillations was performed, but the effect is opposite to that
advocated in our paper: instead of amplification the anharmonicity leads to a
damping of particle production. However, in this paper the form of the inflaton
oscillations is different from ours. It shows that the shape of the signal is indeed
of crucial importance to the efficiency of the particle production. We thank
M.\,Amin for the indication of Ref.~\cite{underwood}. Recently there have appeared a
few more papers~\cite{lozanov, hertzberg-1, hertzberg-2} in which the efficiency of
inflationary heating was studied. However, the mechanism considered there is
different from ours.

\acknowledgments

AD and AS acknowledge the support of the grant of the Russian Federation government
11.G34.31.0047.

\end{document}